\documentclass[10pt,twocolumn,letterpaper]{article}

\usepackage{cvpr}[cvprfinal]      %
\usepackage[accsupp]{axessibility}

\usepackage{graphicx}
\usepackage{amsmath}
\usepackage{amssymb}
\usepackage{amsfonts}
\usepackage{booktabs}
\usepackage{xcolor}
\usepackage{float}
\usepackage{lipsum}
\usepackage{algorithm}
\usepackage{algpseudocode}
\usepackage{comment}
\usepackage{tabularray}
\definecolor{brightlavender}{rgb}{0.75, 0.58, 0.89}
\usepackage{array}
\usepackage{adjustbox}
\usepackage{makecell}
\usepackage{fontawesome5}

\definecolor{cvprblue}{rgb}{0.21,0.49,0.74}
\usepackage[pagebackref,breaklinks,colorlinks,allcolors=cvprblue]{hyperref}
\usepackage[automake, acronym, nopostdot, ]{glossaries}
\makeglossaries
\glsdisablehyper

\newacronym{vfx}{VFX}{visual effects}
\newacronym{sota}{SOTA}{state-of-the-art}
\newacronym{mesa}{MESA}{Modeling of Elevation and Surface Appearance}
\newacronym{pbr}{PBR}{physically based rendering}
\newacronym{osm}{OSM}{OpenStreetMap}
\newacronym{vae}{VAE}{variational autoencoder}
\newacronym{dm}{DM}{diffusion model}
\newacronym{ldm}{LDM}{latent diffusion model}
\newacronym{lora}{LoRA}{low-rank adaptation}
\newacronym{brdf}{BRDF}{bidirectional reflectance distribution function}

\usepackage[capitalize]{cleveref}
\crefname{section}{Sec.}{Secs.}
\Crefname{section}{Section}{Sections}
\Crefname{table}{Table}{Tables}
\crefname{table}{Tab.}{Tabs.}

\def\paperID{18} %
\def\confName{MORSE}
\def\confYear{2025}

\title{MESA: Text-Driven Terrain Generation\\Using Latent Diffusion and Global Copernicus Data}

\author{
    Paul Borne\mbox{-{}-}Pons$^{1,2}$\thanks{First author} \quad
    Mikolaj Czerkawski$^{2,3}$ \quad
    Rosalie Martin$^{1}$ \quad
    Romain Rouffet$^{1}$\\
    $^1$Adobe Research \quad $^2$European Space Agency (ESA) \quad $^3$Asterisk Labs
}

\begin{document}
\maketitle
\begin{abstract}

    Terrain modeling has traditionally relied on procedural techniques, which often require extensive domain expertise and handcrafted rules. In this paper, we present MESA - a novel data-centric alternative by training a diffusion model on global remote sensing data. This approach leverages large-scale geospatial information to generate high-quality terrain samples from text descriptions, showcasing a flexible and scalable solution for terrain generation. The model's capabilities are demonstrated through extensive experiments, highlighting its ability to generate realistic and diverse terrain landscapes. The dataset produced to support this work, the Major TOM Core-DEM extension dataset, is released openly as a comprehensive resource for global terrain data. The results suggest that data-driven models, trained on remote sensing data, can provide a powerful tool for realistic terrain modeling and generation.

\end{abstract}
   
\section{Introduction}
\label{sec:intro}

Terrain is the fundamental component of all 3D outdoor scenes~\cite{Smelik2009ASO,galinReviewDigitalTerrain2019} and consequently, terrain modeling lies at the core of the Video Games and \gls{vfx} industry. It is a complex and time-consuming task, particularly when it involves large-scale landscapes, which are getting more common with the current boom in popularity of open world games. The current \gls{sota} in terrain modeling relies mainly on procedural and simulation methods~\cite{galinReviewDigitalTerrain2019}, which rarely scale well beyond a certain point (compute expensive or lack of realism) and can easily fail to capture the variety of the landscape the world offers.

The recent advances in generative machine learning and especially in the area of diffusion models have paved the way for models that can learn a representation of Earth's landscapes directly from real terrain data. By abstracting the complexity of the underlying physical processes, generative models can learn to reproduce patterns and mutual dependencies between visual features, which can lead to high levels of perceptual realism. This work explores the potential of following a similar data-centric methodology for a joint domain of terrain surface model and optical reflectance.

Consequently, a new type of generative model is proposed here and released, referred to as \gls{mesa}. The model is one of the largest generative models trained on Earth observation data released openly to date, benefiting from global and dense coverage of the open-source Major TOM dataset~\cite{francisMajorTOMExpandable2024}. Furthermore, to enable this work, a global expansion to Major TOM containing Copernicus DEM in 30-metre resolution has been released under open, free access on HuggingFace (\url{https://huggingface.co/datasets/Major-TOM/Core-DEM}).  The model architecture based on Stable Diffusion 2.1 produces diverse and realistic terrains represented by corresponding pairs of optical images and elevation maps from text captions. We demonstrate that the model can generate diverse terrain samples of high quality in response to the text prompt, supporting many shapes of terrain, geographical regions, and times of the year. The inference code can be found at \url{https://github.com/PaulBorneP/MESA} and the model weights at \url{https://huggingface.co/NewtNewt/MESA}.

\begin{figure*}
    \centering
    \includegraphics[width=\linewidth]{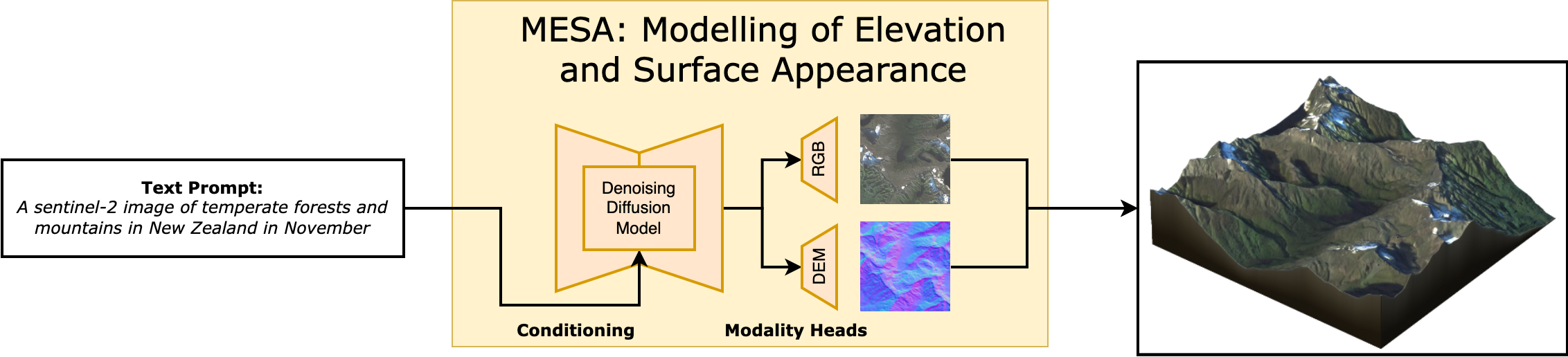}
    \caption{MESA is a novel generative model based on latent denoising diffusion capable of generating 2.5D representations of terrain based on the text prompt conditioning supplied via natural language. The model produces two co-registered modalities of optical and depth maps.}
    \label{fig:dataset_coverage}
\end{figure*}
 
\section{Background}
\label{sec:background}
    The method described in this work lies at the intersection of several domains (terrain modeling, diffusion models, and remote sensing) and the related works for each of them are summarised below.

    \subsection{Terrain Modeling}
    \label{sec:terrain-mod}
        Current terrain modeling revolves mainly around three types of methods (a detailed overview can be found in~\cite{Smelik2009ASO,galinReviewDigitalTerrain2019}): (i) example-based techniques that use real-life terrain as a template to create new ones, (ii) simulations that aim to capture the world complex nature of the world's landscapes with a finite set of equations and heuristic approximations, and (iii) procedural methods that, without relying on real-life examples or strong physical grounding, focus on producing realistic looking outputs with minimal computational cost and direct control mechanisms. Both simulation and procedural methods commonly suffer from the same pitfall, which can be described as the lack of variability in the type of outputs. This is because each type of terrain (mountain, river, dunes) needs to be modeled explicitly by the 3D Artists, and hybrid terrain types are difficult to synthesise. 
        
        Furthermore, procedural methods lose realism when the scale is pushed to extremes, and simulation computational cost often grows exponentially with the terrain dimensions.
        
        While addressing these two issues, example-based methods were originally limited to patch-based algorithms, yielding suboptimal results when compared to the other methods.  More recently, machine learning advances, especially in the field of generative modeling, opened a new line of research, for example, terrain authoring using CGANs~\cite{percheAuthoringTerrainsSpatialised2023} and, more recently, diffusion models~\cite{lochnerInteractiveAuthoringTerrain2023}.
        
        So far these deep learning methods do not tackle the problem of general terrain generation. Instead, they enable the user to author preexisting terrains and also are restricted to authoring of elevation maps.
    
    \subsection{Diffusion Models}
    
        Diffusion models are generative models that aim to approximate a data distribution $p_{\text{Data}}$ from a finite set of samples \cite{sohldickstein2015deepunsupervisedlearningusing, song2021scorebasedgenerativemodelingstochastic}.
        During training, an input image $x \sim p_{\text{Data}}$ is perturbed by adding a gaussian noise $\epsilon \sim \mathcal{N}(0, I)$  to create a noisy image  $x_t = \alpha_t x + \sigma_t \epsilon$. The noise schedule, parameterized by the diffusion time $t$ (where larger values of $t$ correspond to greater noise levels), is defined by its parameters $\alpha_t$ and $\sigma_t$
        
        The diffusion model $\epsilon_\theta$, typically implemented using a UNet or Transformer-based architecture, is trained to denoise $x_t$ by minimizing the score-matching objective. After re-weighting, this objective can be reformulated as an $x$-prediction, $\epsilon$-prediction, or $v$-prediction objective, as described in \cite{kingma2023understandingdiffusionobjectiveselbo}.
        \begin{equation}
            \mathbb{E}_{x \sim p_{\text{Data}}, \epsilon \sim \mathcal{N}(0, I)} \left[ \| y - \epsilon_\theta(x_t; t, c) \|_2^2 \right]
        \end{equation}
        where the target $y$ can be the input noise $\epsilon$, the input image $x$ or the ``velocity'' $v = \alpha_t \epsilon - \sigma_t x$. We can additionally condition the denoising model with side information $c \in \mathbb{R}^D$, which can be a class embedding, text, or other images, etc.  \\ Diffusion models have become particularly popular in image generation due to their ability to produce highly realistic and detailed images, avoiding pitfalls of generative adversarial networks (GANs) such as mode collapse due to its likelihood-based learning objective. \\Latent diffusion models (LDMs) \cite{rombachHighResolutionImageSynthesis2022} first downsample the input $x$ using a \gls{vae} with an encoder $\mathcal{E}$ and a decoder $\mathcal{D}$, such that $\tilde{x} = \mathcal{D}(\mathcal{E}(x))$ is a reconstructed image, perceptually close to the original $x$. Instead of denoising the input image $x$, the diffusion process is used on a downsampled latent representation $z = \mathcal{E}(x)$. This approach reduces computational and memory costs and has formed the basis for the popular Stable Diffusion (SD) model~\cite{rombachHighResolutionImageSynthesis2022}.
        
    \subsection{Diffusion models for joint learning of color and depth maps}
        
        While diffusion models are typically designed to generate standard, 3-channel  RGB images, we aim to simultaneously generate both an RGB image and a corresponding depth map using a single unified model. This approach enables the model to learn not only the visual appearance of scenes but also their underlying 3D structure and the relation between the two. Nevertheless this requires a modification to the standard diffusion model.
        Training a new model from scratch, including the \gls{vae} and the diffusion model (DM), as proposed in some works, can be computationally expensive and may lead to suboptimal results compared to finetuning existing models. Although finetuning the U-Net component of \gls{ldm} is relatively straightforward when employing techniques like \gls{lora}~\cite{hu2021loralowrankadaptationlarge} —finetuning the \gls{vae} is more complex and tedious. The \gls{vae} typically compresses high-frequency information, which remains consistent across different domains (e.g., natural images, satellite images, depth maps), making it viable for encoding and decoding depth maps.
        
        To ensure that the diffusion model generates the required modalities in latent space, different strategies can be adopted. We can either assign different prompts to each modality, as done in \cite{zengRGBLeftrightarrowImage2024} or modify the architecture to include separate output heads for each modality \cite{liu2024hyperhumanhyperrealistichumangeneration}. 
        
        Finally, jointly learning image appearance (color) and geometry (depth map) in a unified network leads to more realistic outputs by incorporating physical information about the image creation process, as demonstrated in~\cite{liu2024hyperhumanhyperrealistichumangeneration}.

    \subsection{Diffusion models for remote sensing generation}
        Diffusion models are now used in various domains of computer vision. As discussed in Section \ref{sec:terrain-mod}, terrain modeling is one such application, and remote sensing is no exception \cite{khannaDiffusionSatGenerativeFoundation2023, sastryGeoSynthContextuallyAwareHighResolution2024, espinosaGenerateYourOwn2023}.
        
        Remote sensing computer vision is inherently challenging due to its divergence from the natural images and its significant semantic diversity, which varies greatly from one geographic region to another. However, it is possible to model this domain with a reasonable amount of compute by fine-tuning pre-trained visual models, such as Stable Diffusion, to balance the effect of under-representation of remote sensing images in general-purpose datasets such as LAION-5B~\cite{czerkawski2023laion5blaioneofilteringbillions}. In \cite{sastryGeoSynthContextuallyAwareHighResolution2024} and \cite{espinosaGenerateYourOwn2023}, the authors explore the use of ControlNets to generate satellite images conditioned on scribbles, Canny edge detection, and OpenStreetMap (OSM) data. This method is computationally efficient since the weights of the UNet and \gls{vae} are frozen, yet it produces photorealistic results. The dense spatial information in the conditioning maps significantly enhances the quality of the generation.
        
        Notably, \cite{khannaDiffusionSatGenerativeFoundation2023} investigates a novel architecture to handle scalar input, such as metadata, and to represent time series. This innovation facilitates conditioned generation tasks, including time-series interpolation.
        
        Although exploring different conditioning methods, all these models require semantically rich inputs, such as textual descriptions or dense spatial maps, to produce satisfactory results. Their primary focus has been on high-resolution (HR) image generation, which aligns more closely with "natural image" features, especially aerial images that have proven to be presented in large image-text datasets~\cite{czerkawski2023laion5blaioneofilteringbillions}. 
        Despite these advancements, the limited availability of high-resolution data constrains the models' ability to achieve a comprehensive representation of Earth's diverse landscapes let alone the fact that training on a dense global dataset of HR images would require unreasonable amount of compute.

\section{Method}
\label{sec:method}

    Building \gls{mesa} relies on a global dataset of reflectance, elevation, and text captions curated specifically to make this work possible and a model based on StableDiffusion but adapted to the 2.5D representation of terrain.

    \subsection{Dataset}

    The dataset used for training the model consists of visual samples from the Major TOM Core dataset, an extension of Major TOM containing Copernicus Digital Elevation Model, and captions used for conditioning the model.
    
    \subsubsection{Major TOM}
        Major TOM (Terrestrial Observation Metaset)~\cite{francisMajorTOMExpandable2024} is an extensible framework for large-scale Earth observation datasets ensuring machine learning readiness and cross usability of each dataset that follows the framework. The structuring idea behind the dataset is to create a regular grid around the Earth and sample images from modalities of interest for each point. Contrary to regular Earth observation data product format in which images are stored in very large tiles 10000 pixels wide, samples from Major Tom Core datasets are smaller and filtered to low cloud coverage, making them immediately applicable in many computer vision deep learning pipelines. The first datasets released within the Major TOM framework already encompass the largest global dataset for Sentinel-1 (S1) and Sentinel-2 (S2) (L2A and L1C). To match this pre-existing dataset, we have curated and released a Major TOM dataset for DEM, Major TOM Core-DEM based on Copernicus DEM (GLO30). Using bilinear interpolation, the tiles originally in WGS84 were resampled to the UTM projection of the desired grid cell before extracting them. We noticed that choosing bilinear interpolation over nearest neighbor interpolation drastically reduced the stairway artifact issue discussed in \cite{galinReviewDigitalTerrain2019} while preserving relatively sharp details. The dataset is released under the Major TOM organisation on HuggingFace with free and open access. Both datasets provide global and dense coverage of Earth's landmass, resulting in a considerable volume of training data indicated in Table~\ref{tab:datasets}
            
        \begin{figure}[h!]
            \centering
            \includegraphics[width=\linewidth]{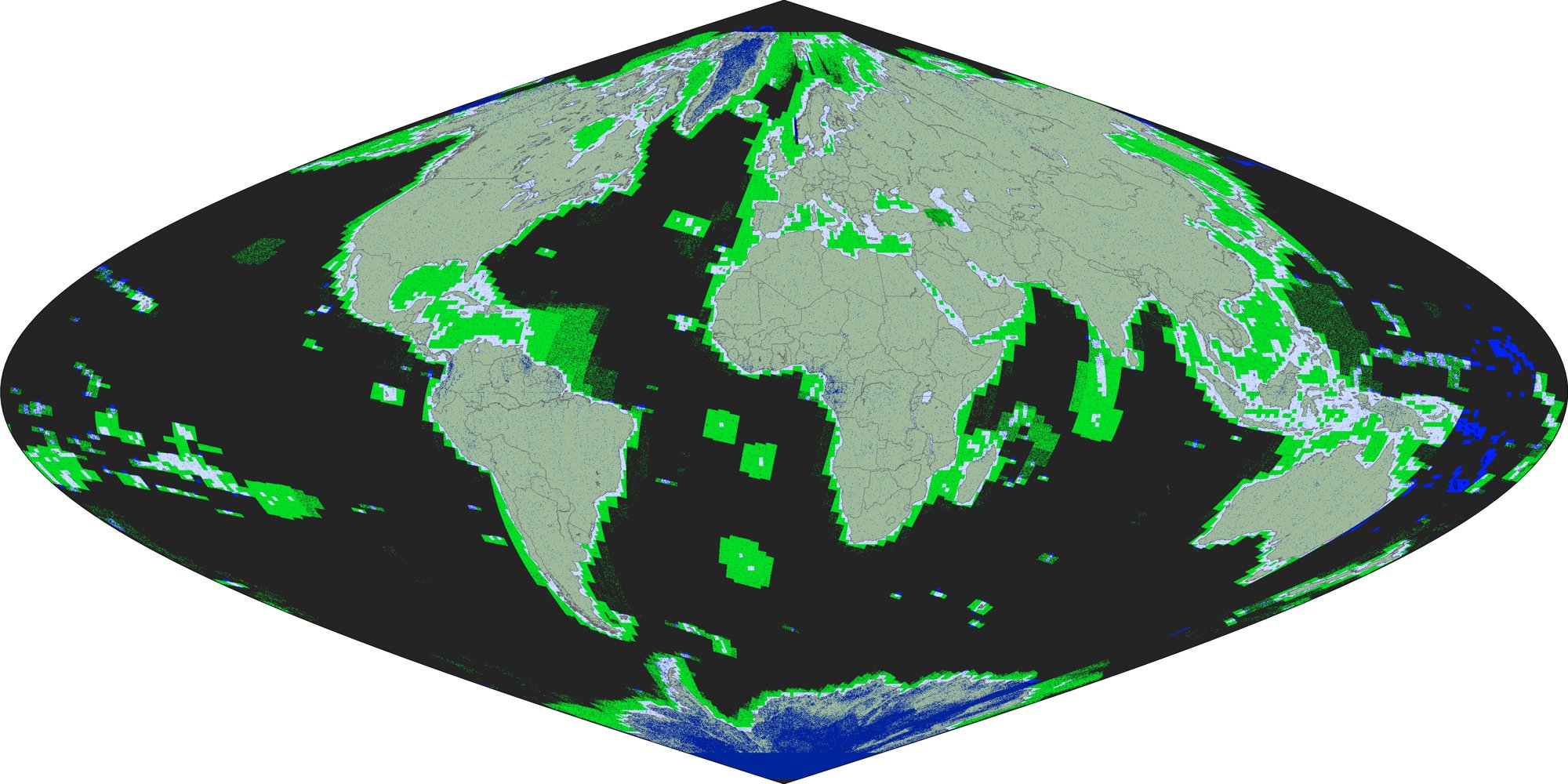}
            \caption{Coverage of the dataset used for this work. Every pixel corresponds to a single cell on the Major TOM grid (10 km). Green marks regions with only Sentinel-2 images available, while blue indicates those with only DEM. Black indicates the absence of any data, while the land and water colors represent the presence of both modalities.}
            \label{fig:dataset_coverage}
        \end{figure}
        
        In addition to the Image and DEM cells, two masks are used:
        \begin{itemize}
            \item a no-data mask that indicates all pixels that have no data in either Image or DEM modality (which can be due to border effects while putting together the respective Major TOM dataset or data corruption in L2A images, especially prominent in remote areas)
            \item a cloud mask available alongside Major TOM core L2A and computed using the SEnSeIv2 algorithm~\cite{Francis_2022,SENSEIv2}
        \end{itemize}
        
        The final mask of valid pixels is obtained through the union of these two masks. 
        
        \subsubsection{Captions from geographical coordinates}\label{sec:GPS caption}
            While text prompts are widely available for natural images with datasets such as LAION-5B \cite{schuhmann2022laion5bopenlargescaledataset}, satellite images typically do not easily come with such captions (other than the heavily biased subset of LAION-5B that contains satellite images contained in CommonCrawl~\cite{czerkawski2023laion5blaioneofilteringbillions}).
            
            The approach to generate global terrain-oriented captions for \gls{mesa} was to use the geographical coordinates of each cell and map dividing the world into regions of interest to create a caption that describes the region in which the cell is located.
            
            \begin{itemize}
                \item A map of countries, following the UN (United Nations) delimitation of borders with this map from the WFP (World Food Program). 
                \item A biome is a distinct geographical region with a specific climate, vegetation, and animal life. It consists of a biological community that has formed in response to its physical environment and regional climate. Following the regions given in \cite{dinersteinEcoregionBasedApproachProtecting2017} we delimit the world in 846 ecoregions that are arranged into 14 types of biomes. While the biome name can give a hint at the topography of the place (An Alpine meadow hints at a mountainous landscape while a plain hints at a flat topography), the biome name mainly gives information on the color and structures that can be seen in the image.
                \item The topographical information can be more easily captured by describing the Earth's geology. Using \cite{namedlandforms} we obtain an indicator that represents the local topology (plain, hills, mountains and so on) and a coarser one that describes the surrounding geological region. This is useful to know the expanse of name entities like "The Alps" or "The Appalachian". During training, we used both the local description and the regional one. 
            \end{itemize}
            
            \begin{table*}[h!]
                \centering
                \caption{Parameters of the Major TOM Core S2-L2A and DEM datasets}
                \resizebox{\linewidth}{!}{
                \begin{tabular}{l l l l l}
                    \toprule
                    \textbf{Source} & \textbf{Modality Type} & \begin{minipage}{2cm} \centering \textbf{Number of} \\ \textbf{Patches} \end{minipage} & \textbf{Patch Size} & \textbf{Total Pixels} \\
                    \midrule
                    Copernicus DEM 30 & Digital Surface Model (DSM) & 1,837,843 & 356 x 356 (30 m) & \textgreater 1.654 Billion \\
                    Sentinel-2 Level-2A & Optical Multispectral & 2,245,886 & $1,068 \times 1,068$ (10 m) & \textgreater 2.564 Trillion \\
                    \bottomrule
                \end{tabular}}
                \label{tab:datasets}
            \end{table*}
            
            To get the final extent of the dataset, all cells are considered where it is possible to acquire a corresponding Image and DEM visible as shown on Figure \ref{fig:dataset_coverage}. After filtering out oceans and large water bodies with flat elevation, antarctic samples (due to poor signal quality), and cells with misaligned projections or missing data, the training dataset amounts to 1.3 million 10$\times$10 km$^2$ 2.5D terrains which is equivalent to most of Earth's land surface beyond Antarctica.
    
    \subsection{Terrain Generation}
        
        To create our 2.5D terrains, we adapt the Stable Diffusion 2.1 architecture to make room for the additional depth channel and fine-tune the resulting model on the triplet of RGB, DEM, and text. To adjust to the original architecture, each dataset image $x_{I}\in \mathbb{R}^{C\times H \times W}$ is cropped to $H=W=768$ and $C=3$. Our DEM cells are notably 3 times coarser than their image counterpart and only require one channel. The initial experiments attempted to take advantage of the coarse spatial resolution of the latent space being close to DEM native resolution and to learn the DEM modality directly as an extra latent channel concatenated to the base. However, this approach did not seem effective and the final design reshapes the DEM so that $x_{D}\in \mathbb{R}^{3\times H \times W}$ and encode it with the same frozen \gls{vae} used for $\mathbf{x}_{I}$. This results in two latent vectors $\mathbf{z}_{I}=\mathcal{E}(\mathbf{x}_I)$ and $,\mathbf{z}_{D}=\mathcal{E}(\mathbf{x}_D) \in \mathbb{R}^{ C'\times H' \times\ W'}$ with $C'=4$ and $H'=W'=96$. Subsequently, the mask is also resized to the dimension of the latent space. The resized mask is represented as $\mathbf{z}_{M}$ for uniformity of notation.
        
        Following an approach similar to previous works~\cite{liu2024hyperhumanhyperrealistichumangeneration}, the model is trained to simultaneously
        denoise the depth and the RGB image using two separate modality-specific heads. While the two heads makes the model expressive enough to model the specificity of each modality, the common backbone ensures the spatial alignment and semantic coherence of the modalities (ie. shadows next to hilly areas).
        We capture the joint distribution of image and DEM with a single U-Net model by simultaneously denoising the different modalities, which can be trained with a simple v-prediction objective. Since the aim is to train a model to generate data without the artifacts due to clouds or missing data, pixels affected by either of these issues are masked out using $\mathbf{z}_{M}$. The final notation is given by
        \begin{equation}   
            \mathcal{L}^{v\text{-pred}} = 
            \mathbb{E}_{\mathbf{z}_{I,D}, \epsilon,t,c,\mathbf{z}_{M}} \Biggl[
            \Bigl\|\mathcal{I}_{\mathbf{z}_{M}}( \underbrace{\epsilon_{\theta} ( \mathbf{v}_{I,D,t}, c)}_{\mathbf{\hat{v}}_{I,D,t}} - \mathbf{v}_{I,D,t})\Bigr\|_2^2 \Biggr]
        \end{equation}
        
        $\mathbf{z}_{I,D}=(\mathbf{z}_{I},\mathbf{z}_{D}) \sim p_{rgbd}$  the joint distribution of latent images and DEMs, $\mathbf{x}_M$ the resized mask, and $\mathcal{I}$ the indicator function.
        
        $\mathbf{v}_{t_I} = \alpha_{t_I}\epsilon_I - \sigma_{t_I}\mathbf{z_I}  $, and $\mathbf{v}_{t_D} = \alpha_{t_D}\epsilon_D - \sigma_{t_D} \mathbf{z_D}$  the target velocities. Subsequently $\epsilon_{\theta} ((\mathbf{v}_{t_I},\mathbf{v}_{t_D}), c)= (\mathbf{\hat{v}}_{t_I},\mathbf{\hat{v}}_{t_D})$ with $\mathbf{\hat{v}}_{t_I} = \alpha_{t_I}\epsilon_I - \sigma_{t_I}\mathbf{\hat{z}_I} $, and $\mathbf{\hat{v}}_{t_D} = \alpha_{t_D}\epsilon_D - \sigma_{t_D} \mathbf{\hat{z}_D}$  the predicted velocities and $\mathbf{\hat{z}}$ being the result of one step denoising. For ease of notation we write $\mathbf{v}_{I,D,t}=(\mathbf{v}_{t_I},\mathbf{v}_{t_D})$ and $\mathbf{\hat{v}}_{I,D,t}=(\mathbf{\hat{v}}_{t_I},\mathbf{\hat{v}}_{t_D})$

        $\epsilon_I$ and $\epsilon_D \sim \mathcal{N}(0, \mathbf{I})$ are two independent samples of Gaussian noise for the RGB and DEM, respectively.
        
        $t_I$, $t_D \sim \mathcal{U}[1, T]$ are the sampled time-steps that control the scale of added Gaussian noise. We could use different time-steps for each modality, but as mentioned in \cite{liu2024hyperhumanhyperrealistichumangeneration}, such a strategy only gives $10^{-6}$ probability to sample each perturbation state (given $T = 1000$), which may hinder convergence. Therefore we use $t=t_x=t_D$.
        
        Finally, the caption $c \in \mathbb{R}^L$ with $L$ being the output dimension of the CLIP text encoder~\cite{radford2021learningtransferablevisualmodels} is used to compute a vector representation of our caption. Using descriptors introduced in Section \ref{sec:GPS caption} our prompts follow the structure "A Sentinel-2 image of \{Biome\} and \{Geological\} in \{Country\} in \{Month\}". To ensure that the model does not merely respond to prompts following that exact structure we randomly drop one, two, or three of these descriptors. We also randomly anonymize the ecoregion name and shuffle between local and regional geological descriptors.
        
        To generate a 2.5D terrain, two independent samples of Gaussian noise are drawn as $\epsilon_I$ and $\epsilon_D \sim \mathcal{N}(0, \mathbf{I})$. Using DDIM scheduler and the Trained U-Net conditioned by the caption $c$ we generate the two latent vectors $(\mathbf{\hat{z}}_{I},\mathbf{\hat{z}}_{D})$ that we both pass through the frozen \gls{vae} decoder to get the final generated Image and DEM $
        \mathbf{\hat{x}}_{I}=\mathcal{D}(\mathbf{\hat{z}}_{I})$
        $
        \mathbf{\hat{x}}_{D}=\mathcal{D}(\mathbf{\hat{z}}_{D})$
        
        \begin{figure}
          \centering
          \includegraphics[width=\linewidth]{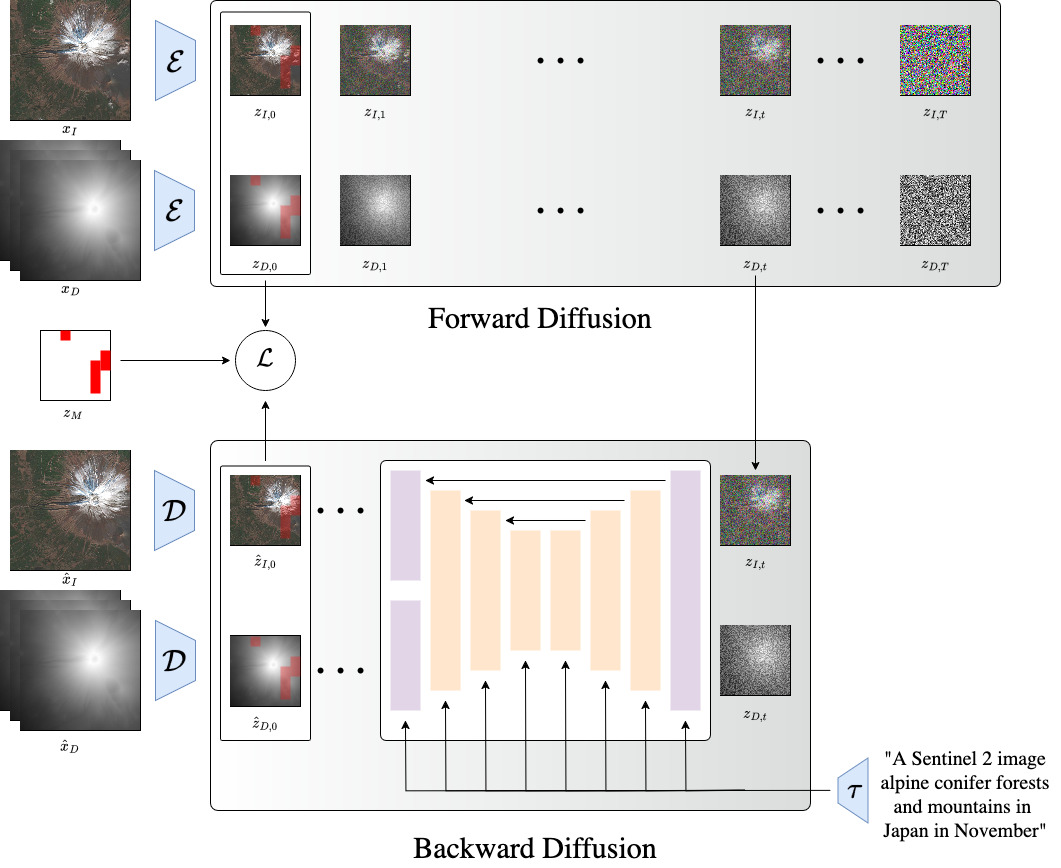}
           \caption{Using Stable Diffusion 2.1 weights, we project RGB and depth maps into latent space with \textcolor{blue}{frozen VAE encoders}. Latents are noised and denoised conditionally on captions via a modified U-Net. We \textcolor{red}{mask the loss} with $\mathbf{z}_M$ to focus on cloud-free pixels, enabling cloud-free terrain generation.}
           \label{fig:architecture}
        \end{figure}
        
        An overview of the architecture can be seen in Figure \ref{fig:architecture}. Using Stable Diffusion 2.1 weights, the \textcolor{blue}{frozen image encoders/decoders (\gls{vae})} are used to project both RGB and depth maps (concatenated 3 times to match the 3-channel input) to the latent space. During training, both latents are noised (forward diffusion) with independent noise and then jointly denoised conditionally on caption $c$ (embedded using the \textcolor{blue}{frozen text encoder}) using a modified U-Net.

            \begin{table*}%
        \centering     
        \caption{Variability and influence of seed. \gls{mesa} can produce many diverse samples for any caption, while preserving the same terrain characteristics.}
        \renewcommand{\arraystretch}{1.5} %
        \setlength{\tabcolsep}{4pt} %
        \begin{tabular}{>{\centering\arraybackslash}m{4cm}>{\centering\arraybackslash}m{2cm}>{\centering\arraybackslash}m{2cm}>{\centering\arraybackslash}m{2cm}>{\centering\arraybackslash}m{2cm}>{\centering\arraybackslash}m{2cm}>{\centering\arraybackslash}m{2cm}}
        \cline{2-7}
        & \multicolumn{2}{c}{\textbf{seed 42}} & \multicolumn{2}{c}{\textbf{seed 50}} & \multicolumn{2}{c}{\textbf{seed 666}} \\
        \hline
        Caption & RGB & DEM & RGB & DEM & RGB & DEM \\ 
        \hline
        "broadleaf forests and hills in Germany in August" & 
        \vspace{5pt}
        \adjustbox{max width=2cm}{\includegraphics{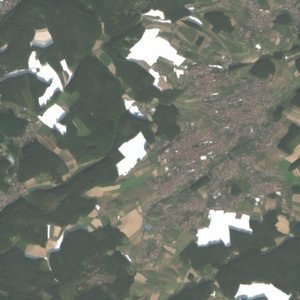}} & 
        \vspace{5pt}
        \adjustbox{max width=2cm}{\includegraphics{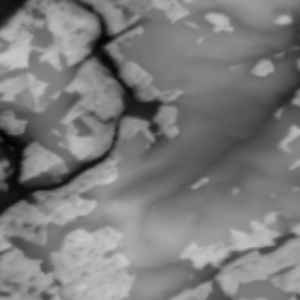}} &
        \vspace{5pt}
        \adjustbox{max width=2cm}{\includegraphics{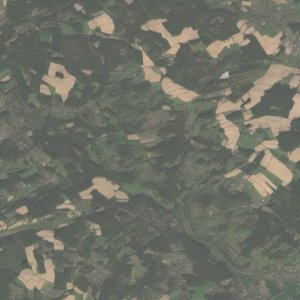}} & 
        \vspace{5pt}
        \adjustbox{max width=2cm}{\includegraphics{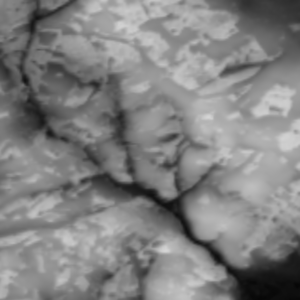}} &
        \vspace{5pt}
        \adjustbox{max width=2cm}{\includegraphics{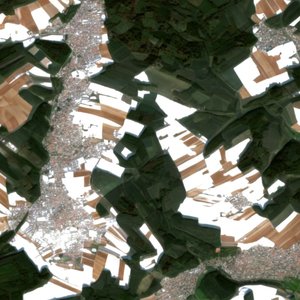}} & 
        \vspace{5pt}
        \adjustbox{max width=2cm}{\includegraphics{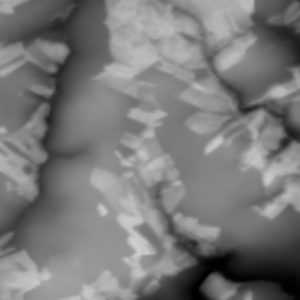}} \\
        \hline
        "dry woodlands steppe and mountains in Tunisia in July" & 
        \vspace{5pt}
        \adjustbox{max width=2cm}{\includegraphics{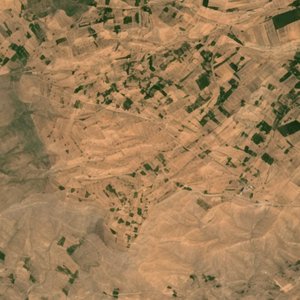}} & 
        \vspace{5pt}
        \adjustbox{max width=2cm}{\includegraphics{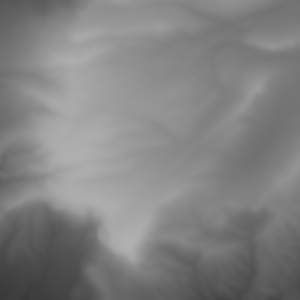}} &
        \vspace{5pt}
        \adjustbox{max width=2cm}{\includegraphics{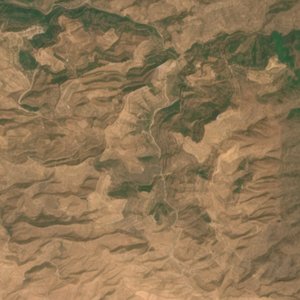}} & 
        \vspace{5pt}
        \adjustbox{max width=2cm}{\includegraphics{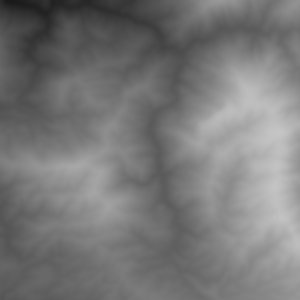}} &
        \vspace{5pt}
        \adjustbox{max width=2cm}{\includegraphics{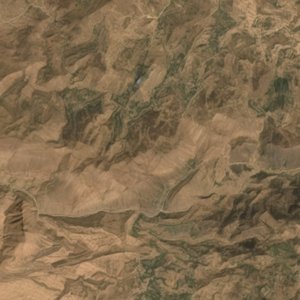}} & 
        \vspace{5pt}
        \adjustbox{max width=2cm}{\includegraphics{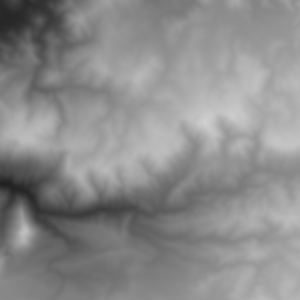}} \\
        \hline
        "tundra and mountains in United States of America in November" & 
        \vspace{5pt}
        \adjustbox{max width=2cm}{\includegraphics{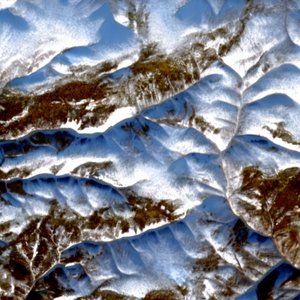}} & 
        \vspace{5pt}
        \adjustbox{max width=2cm}{\includegraphics{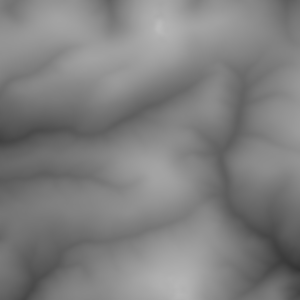}} &
        \vspace{5pt}
        \adjustbox{max width=2cm}{\includegraphics{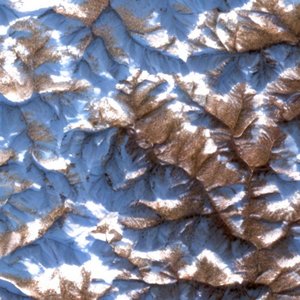}} & 
        \vspace{5pt}
        \adjustbox{max width=2cm}{\includegraphics{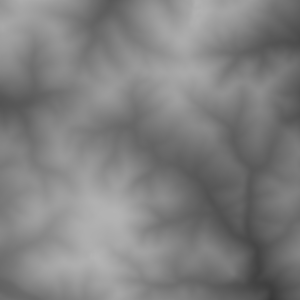}} &
        \vspace{5pt}
        \adjustbox{max width=2cm}{\includegraphics{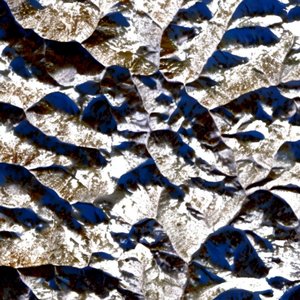}} & 
        \vspace{5pt}
        \adjustbox{max width=2cm}{\includegraphics{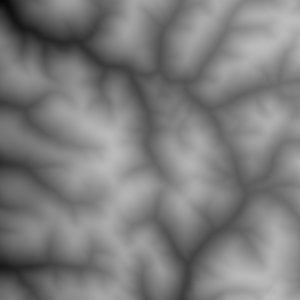}} \\
        
        \hline
        "moist forests and plains in Bolivia in July" & 
        \vspace{5pt}
        \adjustbox{max width=2cm}{\includegraphics{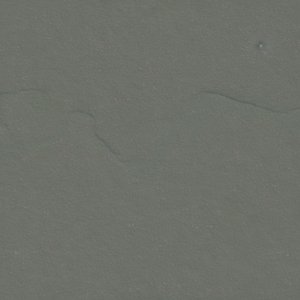}} & 
        \vspace{5pt}
        \adjustbox{max width=2cm}{\includegraphics{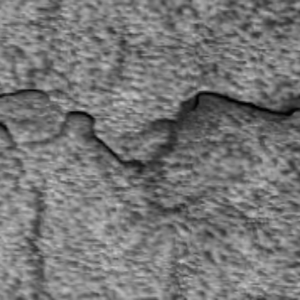}} &
        \vspace{5pt}
        \adjustbox{max width=2cm}{\includegraphics{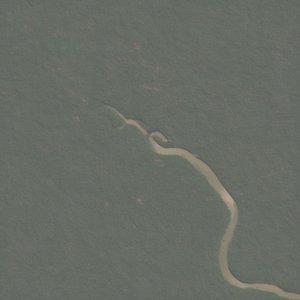}} & 
        \vspace{5pt}
        \adjustbox{max width=2cm}{\includegraphics{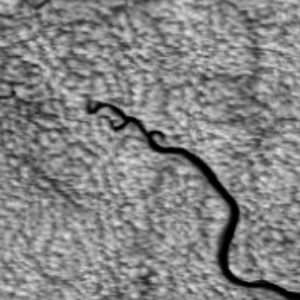}} &
        \vspace{5pt}
        \adjustbox{max width=2cm}{\includegraphics{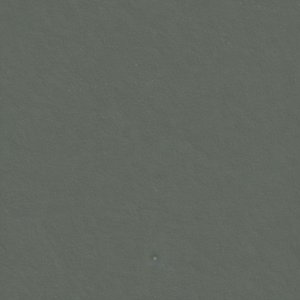}} & 
        \vspace{5pt}
        \adjustbox{max width=2cm}{\includegraphics{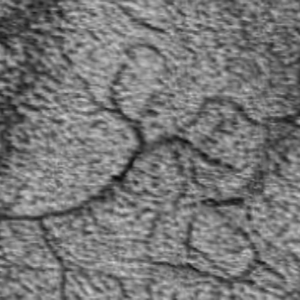}} \\
        \hline
        \end{tabular}
        \label{tab:exp_seed}
    \end{table*}

        The diffusion U-Net architecture consists of down-sampling, middle, and up-sampling blocks, constructed from convolutional layers and self-/cross-attention mechanisms. Specifically, the DownBlocks reduce the resolution of the input noisy latent to generate lower-resolution hidden states, while the UpBlocks perform the reverse, progressively upscaling features to predict the noise. Inspired by \cite{liu2024hyperhumanhyperrealistichumangeneration}, we initially considered duplicating the \textcolor{brightlavender}{first DownBlock} and  \textcolor{brightlavender}{last  UpBlock} to create separate heads for each modality, each initialized with Stable Diffusion 2.1 weights. However, we found that using a shared \textcolor{brightlavender}{first DownBlock} for all modalities yields comparable results while reducing parameter count and simplifying the architecture. The output of this shared head is averaged across modalities and then passed to the \textcolor{orange}{backbone}, whose output is then distributed to the corresponding output heads (\textcolor{brightlavender}{last UpBlocks}) for each modality.
       
        Using the $\mathbf{z}_M$, we \textcolor{red}{mask the loss} $\mathcal{L}$ and hence only backpropagate through valid and cloud-free pixels. Assuming that structure is preserved when projecting the input images into the latent space, it enables us to learn a cloud-free distribution of images.

        The model is trained on 8 NVIDIA A100 GPUs, with a batch size of 128, for 80,000 iterations at a constant learning rate of $10^{-5}$ of an AdamW optimizer. For all experiments, we use DDIM sampling with 50 steps and a guidance scale of 7~\cite{songDenoisingDiffusionImplicit2022}.

\section{Experiments}
\label{sec:experiments}

    The experimental section involves primarily qualitative results to demonstrate the quality of the synthesis by \gls{mesa}. There are no existing benchmarks for text-based terrain modeling, and it is hoped that this work can act as a first step in that direction.

    \subsection{Variablity and  influence of seed}

        The first set of experiments demonstrates the diversity of generated data. Table~\ref{tab:exp_seed} shows multiple generations produced for different seeds (columns) with a fixed caption (rows). Depending on seed, the model can generate a diverse set of samples, all exhibiting similar characteristics that reflect the conditions specified in the caption.

    \subsection{Influence of captions}

            \begin{table*}%
        \centering
        \caption{Influence of Captions. Named prompts indicates prompts with context of generating specific named are (such as Chihuahua). The remaining rows contain standard prompt with various parts dropped out, according to the legend: \textcolor{black}{\faIcon{tree}} - biome, \textcolor{black}{\faIcon{mountain}} - geological, \textcolor{black}{\faIcon{flag}} - country, \textcolor{black}{\faIcon{calendar-alt}} - month}
        \renewcommand{\arraystretch}{1.5} %
        \setlength{\tabcolsep}{4pt} %
        \begin{tabular}{>{\centering\arraybackslash}m{4cm}>{\centering\arraybackslash}m{2cm}>{\centering\arraybackslash}m{2cm}>{\centering\arraybackslash}m{2cm}>{\centering\arraybackslash}m{2cm}>{\centering\arraybackslash}m{2cm}>{\centering\arraybackslash}m{2cm}}
        \cline{2-7}
        & \multicolumn{2}{c}{\textbf{cell 1}} & \multicolumn{2}{c}{\textbf{cell 2}} & \multicolumn{2}{c}{\textbf{cell 3}} \\
        \hline
        &\multicolumn{2}{c}{\adjustbox{max width=4cm}{\parbox{4cm}{mixed forests (\textcolor{black}{\faIcon{tree}}) and hills (\textcolor{black}{\faIcon{mountain}}) in France (\textcolor{black}{\faIcon{flag}}) in September (\textcolor{black}{\faIcon{calendar-alt}})}}} & 
        \multicolumn{2}{c}{\adjustbox{max width=4cm}{\parbox{4cm}{desert (\textcolor{black}{\faIcon{tree}}) and mountains (\textcolor{black}{\faIcon{mountain}}) in United States of America (\textcolor{black}{\faIcon{flag}}) in May (\textcolor{black}{\faIcon{calendar-alt}})}}} & 
        \multicolumn{2}{c}{\adjustbox{max width=4cm}{\parbox{4cm}{conifer deciduous forests (\textcolor{black}{\faIcon{tree}}) and mountains (\textcolor{black}{\faIcon{mountain}}) in Turkey (\textcolor{black}{\faIcon{flag}}) in January (\textcolor{black}{\faIcon{calendar-alt}})}}}\\
        \textcolor{green}{\faIcon{tree}} \textcolor{green}{\faIcon{mountain}} \textcolor{green}{\faIcon{flag}} \textcolor{green}{\faIcon{calendar-alt}} & 
        \vspace{5pt}
        \adjustbox{max width=2cm}{\includegraphics{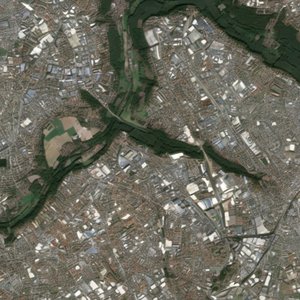}} & 
        \vspace{5pt}
        \adjustbox{max width=2cm}{\includegraphics{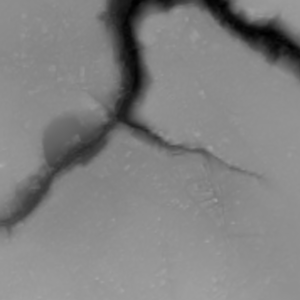}} &
        \vspace{5pt}
        \adjustbox{max width=2cm}{\includegraphics{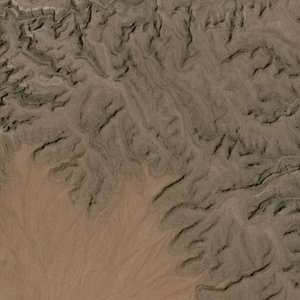}} & 
        \vspace{5pt}
        \adjustbox{max width=2cm}{\includegraphics{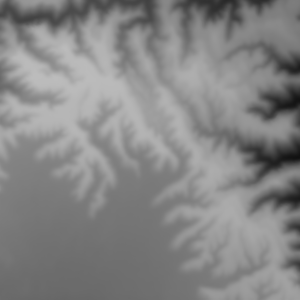}} &
        \vspace{5pt}
        \adjustbox{max width=2cm}{\includegraphics{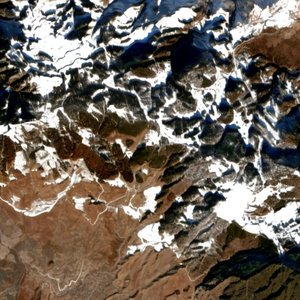}} & 
        \vspace{5pt}
        \adjustbox{max width=2cm}{\includegraphics{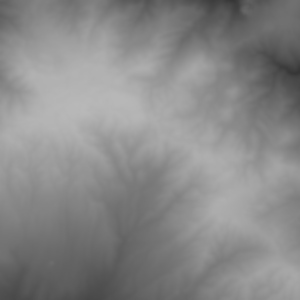}} \\
        \hline
        
        \textcolor{green}{\faIcon{tree}} \textcolor{green}{\faIcon{mountain}} \textcolor{green}{\faIcon{flag}} \textcolor{red}{\faIcon{calendar-alt}} & 
        \vspace{5pt}
        \adjustbox{max width=2cm}{\includegraphics{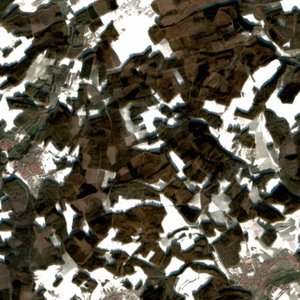}} & 
        \vspace{5pt}
        \adjustbox{max width=2cm}{\includegraphics{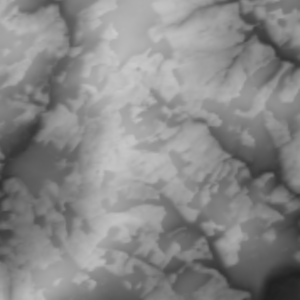}} &
        \vspace{5pt}
        \adjustbox{max width=2cm}{\includegraphics{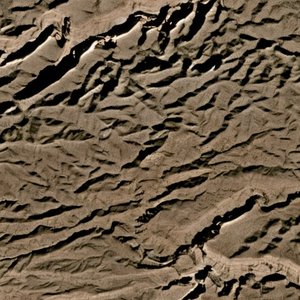}} & 
        \vspace{5pt}
        \adjustbox{max width=2cm}{\includegraphics{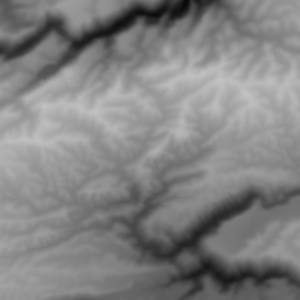}} &
        \vspace{5pt}
        \adjustbox{max width=2cm}{\includegraphics{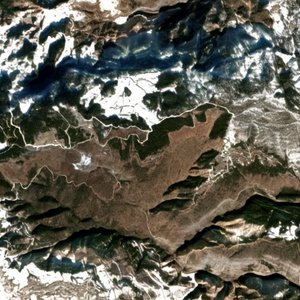}} & 
        \vspace{5pt}
        \adjustbox{max width=2cm}{\includegraphics{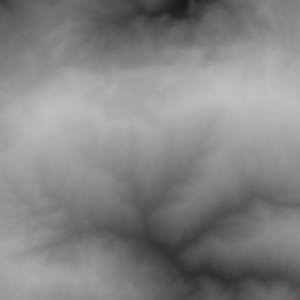}} \\
        \hline
        
        \textcolor{red}{\faIcon{tree}} \textcolor{green}{\faIcon{mountain}} \textcolor{green}{\faIcon{flag}} \textcolor{red}{\faIcon{calendar-alt}} & 
        \vspace{5pt}
        \adjustbox{max width=2cm}{\includegraphics{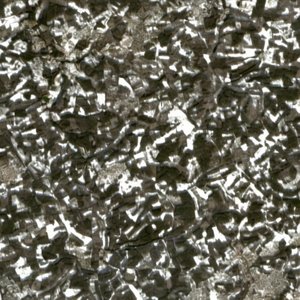}} & 
        \vspace{5pt}
        \adjustbox{max width=2cm}{\includegraphics{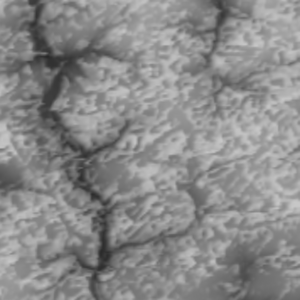}} &
        \vspace{5pt}
        \adjustbox{max width=2cm}{\includegraphics{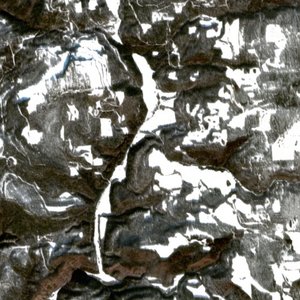}} & 
        \vspace{5pt}
        \adjustbox{max width=2cm}{\includegraphics{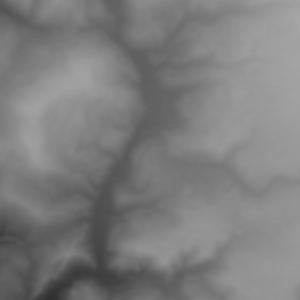}} &
        \vspace{5pt}
        \adjustbox{max width=2cm}{\includegraphics{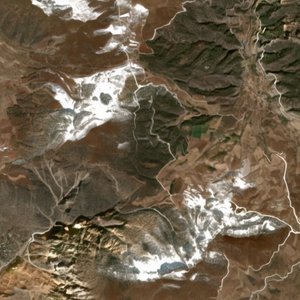}} & 
        \vspace{5pt}
        \adjustbox{max width=2cm}{\includegraphics{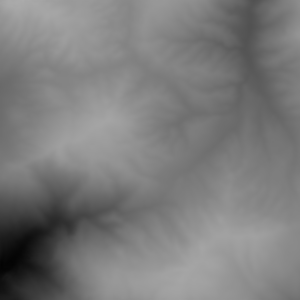}} \\
        \hline
        
        \textcolor{green}{\faIcon{tree}} \textcolor{red}{\faIcon{mountain}} \textcolor{green}{\faIcon{flag}} \textcolor{red}{\faIcon{calendar-alt}} & 
        \vspace{5pt}
        \adjustbox{max width=2cm}{\includegraphics{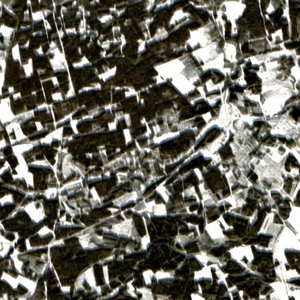}} & 
        \vspace{5pt}
        \adjustbox{max width=2cm}{\includegraphics{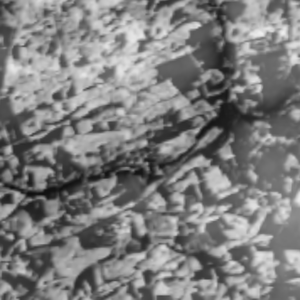}} &
        \vspace{5pt}
        \adjustbox{max width=2cm}{\includegraphics{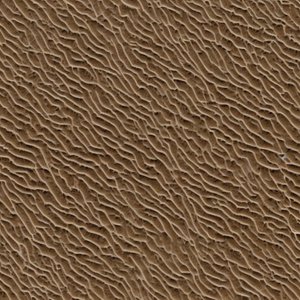}} & 
        \vspace{5pt}
        \adjustbox{max width=2cm}{\includegraphics{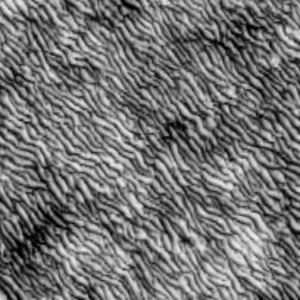}} &
        \vspace{5pt}
        \adjustbox{max width=2cm}{\includegraphics{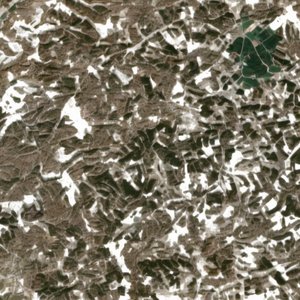}} & 
        \vspace{5pt}
        \adjustbox{max width=2cm}{\includegraphics{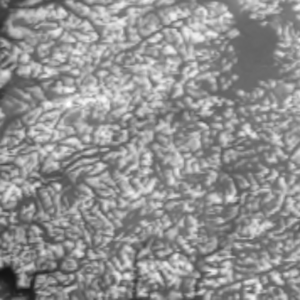}} \\
        \hline
        
        \textcolor{green}{\faIcon{tree}} \textcolor{green}{\faIcon{mountain}} \textcolor{red}{\faIcon{flag}} \textcolor{red}{\faIcon{calendar-alt}} & 
        \vspace{5pt}
        \adjustbox{max width=2cm}{\includegraphics{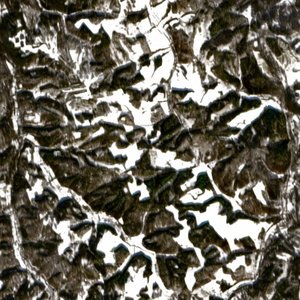}} & 
        \vspace{5pt}
        \adjustbox{max width=2cm}{\includegraphics{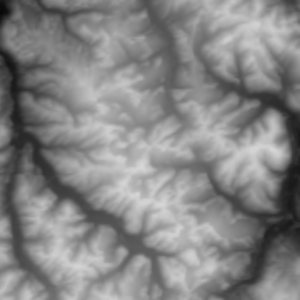}} &
        \vspace{5pt}
        \adjustbox{max width=2cm}{\includegraphics{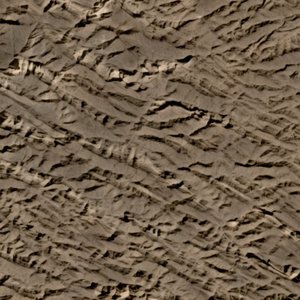}} & 
        \vspace{5pt}
        \adjustbox{max width=2cm}{\includegraphics{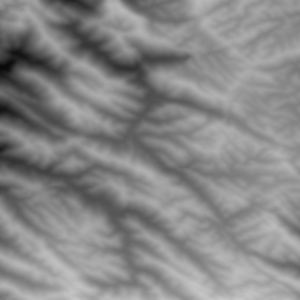}} &
        \vspace{5pt}
        \adjustbox{max width=2cm}{\includegraphics{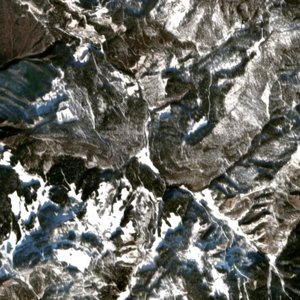}} & 
        \vspace{5pt}
        \adjustbox{max width=2cm}{\includegraphics{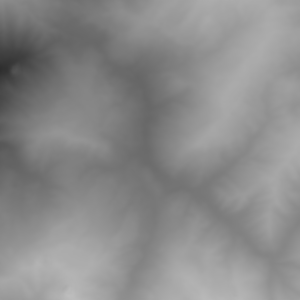}} \\
        \hline
        
        \textcolor{red}{\faIcon{tree}} \textcolor{red}{\faIcon{mountain}} \textcolor{green}{\faIcon{flag}} \textcolor{red}{\faIcon{calendar-alt}} & 
        \vspace{5pt}
        \adjustbox{max width=2cm}{\includegraphics{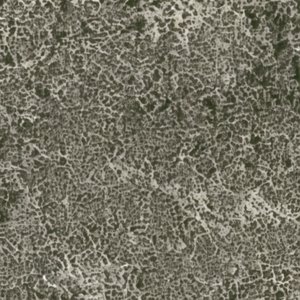}} & 
        \vspace{5pt}
        \adjustbox{max width=2cm}{\includegraphics{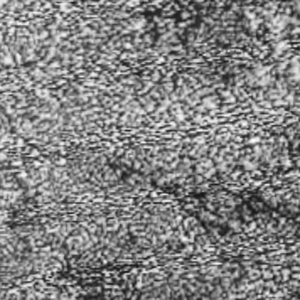}} &
        \vspace{5pt}
        \adjustbox{max width=2cm}{\includegraphics{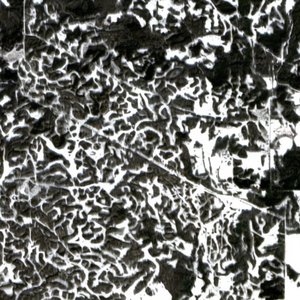}} & 
        \vspace{5pt}
        \adjustbox{max width=2cm}{\includegraphics{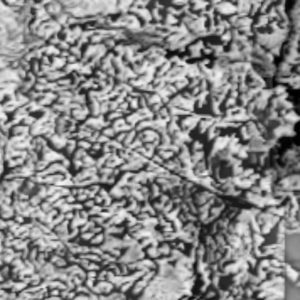}} &
        \vspace{5pt}
        \adjustbox{max width=2cm}{\includegraphics{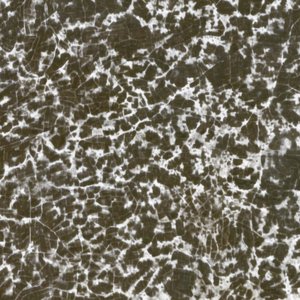}} & 
        \vspace{5pt}
        \adjustbox{max width=2cm}{\includegraphics{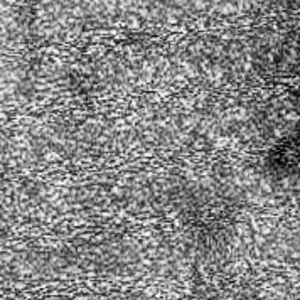}} \\
    
        \hline
        \end{tabular}
        \label{tab:exp_cap}
    \end{table*}

        Another important experiment to conduct is to test how the model responds to different information provided in captions. \Gls{mesa} can support four types of descriptors in the text prompt: biome, geological, country, and month of the year. Table~\ref{tab:exp_cap} demonstrates several configurations for information contained in the prompt (rows) for several different samples (columns). The first row contains a prompt with all four descriptors present (indicated by green icons), while the following rows test for the absence of particular pieces of the full prompt. First of all, biome information is quite important to include to control both features present in the image as well as their scale. The geological has a strong influence over the share of the elevation, and without it, the samples may look quite unrealistic. The date (only used in the first row), helps to generate an appropriate state of the flora and weather conditions (such as preventing snow from being generated) but can still be inconsistent in some cases. The exclusion of country information does not have an effect as severe, and as shown in the third last row, including the biome and geological information in the prompt is enough to generate realistic samples. This is also demonstrated in the row where only country information is provided, and the samples look quite unrealistic, leading to a very texture-like sample appearance.

    \subsection{Influence of shadow correction}

        Finally, the influence of the shadow correction technique is tested by comparing two models, one trained on original L2A data and one trained on shadow-corrected L2A variant.

        The shadow correction is carried out via histogram matching between L1C image by transferring the colour distribution of the L2A image onto it.
        
        Table~\ref{tab:exp_shd} contains a set of results for two independently trained variants of \gls{mesa}. The amount of artifacts due to the L2A processing of Sentinel-2 data is reduced, while generating terrains with similar characteristics.

        \begin{table*}%
        \centering
        \caption{Influence of shadow correction for two independently trained versions of \gls{mesa}. The model trained on shadow-corrected data exhibits superior visual quality due to the fewer artefacts present in the training data.}
        \renewcommand{\arraystretch}{1.5} %
        \setlength{\tabcolsep}{4pt} %
        \begin{tabular}{>{\centering\arraybackslash}m{4cm}>{\centering\arraybackslash}m{3cm}>{\centering\arraybackslash}m{3cm}>{\centering\arraybackslash}m{3cm}>{\centering\arraybackslash}m{3cm}}
        \cline{2-5}
        &\multicolumn{2}{c}{\textbf{w/ shadow correction}} & \multicolumn{2}{c}{\textbf{wo/ shadow correction}}\\
        \hline
         Caption & RGB & DEM & RGB & DEM  \\ 
        \hline
        "temperate forests and mountains in New Zealand in March" & 
        \vspace{5pt}
        \adjustbox{max width=3cm}{\includegraphics{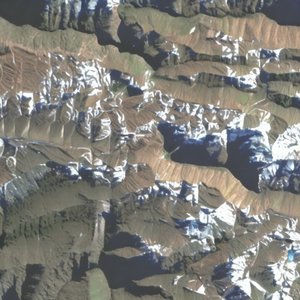}} & 
        \vspace{5pt}
        \adjustbox{max width=3cm}{\includegraphics{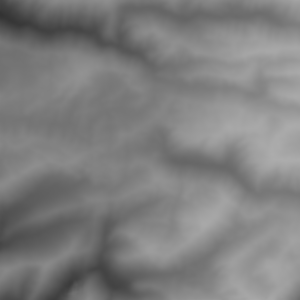}} &
        \vspace{5pt}
        \adjustbox{max width=3cm}{\includegraphics{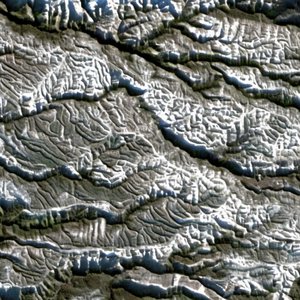}} & 
        \vspace{5pt}
        \adjustbox{max width=3cm}{\includegraphics{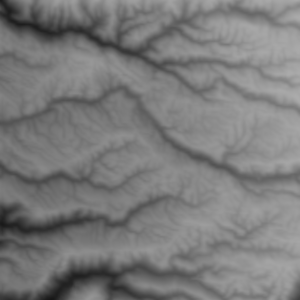}} \\
        \hline
        "temperate forests and mountains in Chile in March" & 
        \vspace{5pt}
        \adjustbox{max width=3cm}{\includegraphics{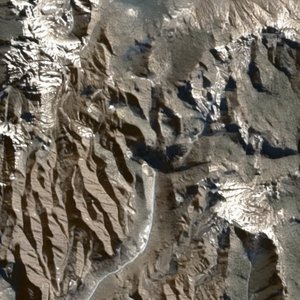}} & 
        \vspace{5pt}
        \adjustbox{max width=3cm}{\includegraphics{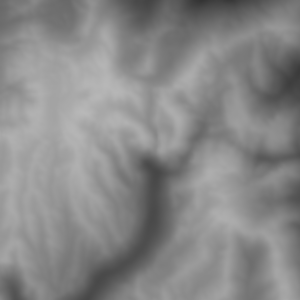}} &
        \vspace{5pt}
        \adjustbox{max width=3cm}{\includegraphics{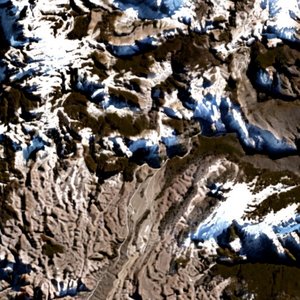}} & 
        \vspace{5pt}
        \adjustbox{max width=3cm}{\includegraphics{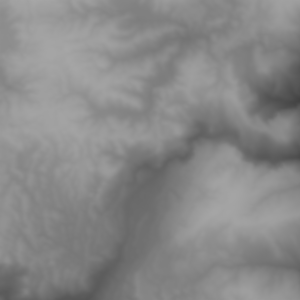}} \\
        
        \hline
        "moist forests andplains in Brazil in May" & 
        \vspace{5pt}
        \adjustbox{max width=3cm}{\includegraphics{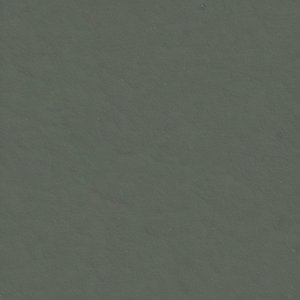}} & 
        \vspace{5pt}
        \adjustbox{max width=3cm}{\includegraphics{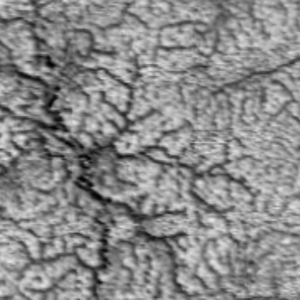}} &
        \vspace{5pt}
        \adjustbox{max width=3cm}{\includegraphics{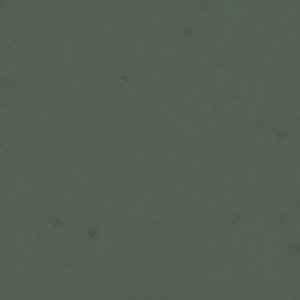}} & 
        \vspace{5pt}
        \adjustbox{max width=3cm}{\includegraphics{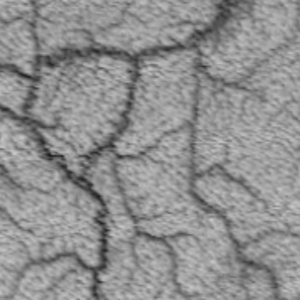}} \\
        \hline
        "montane rain forests and mountains in Indonesia in July" & 
        \vspace{5pt}
        \adjustbox{max width=3cm}{\includegraphics{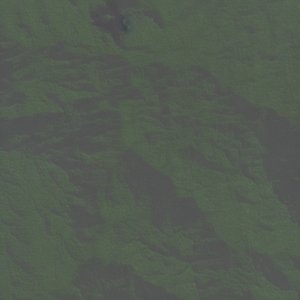}} & 
        \vspace{5pt}
        \adjustbox{max width=3cm}{\includegraphics{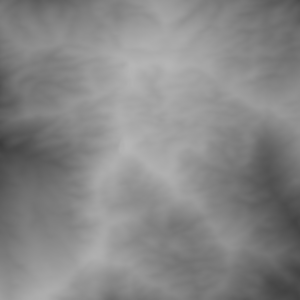}} &
        \vspace{5pt}
        \adjustbox{max width=3cm}{\includegraphics{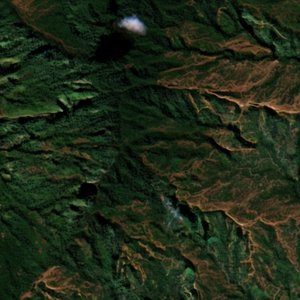}} & 
        \vspace{5pt}
        \adjustbox{max width=3cm}{\includegraphics{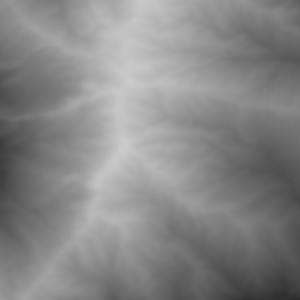}} \\
        \hline
        \end{tabular}
        
        \label{tab:exp_shd}
    \end{table*}

\section{Conclusion}
\label{sec:conclusion}

    This work introduces \gls{mesa}, a text-driven diffusion model for generating diverse terrains based on features specified using natural language. The model has been trained on a global and dense coverage of Major TOM dataset (minus 10 percent sampled randomly and kept for testing), allowing to account for all kinds of terrain present currently on Earth. This indicates a unique value of global and open remote sensing data from the Copernicus programme.

    The model is capable of generating 2.5D representations (optical and depth) of diverse terrains in response to text prompts that control the biome, geological features, country context, and season. This has been demonstrated through a series of qualitative experiments, and the model exhibits promising performance
    to a degree in which it could support creative pipelines where terrain modeling is needed.

{\small
    \bibliographystyle{ieee_fullname}
    \bibliography{main}
}

\end{document}